\documentstyle[12pt]{article}

\def \be{\begin{equation}}
\def \ee{\end{equation}}
\def \bea{\begin{eqnarray}}
\def \eea{\end{eqnarray}}
\begin{document}
{\hskip 1cm  }
\\
\begin{center}
{\Large{\bf{ The Instanton Solution of Forced Burgers Equation in 
Polyakov's Approach}}}
\vskip .5cm   
{\large{ M. R. Rahimi Tabar}}
\vskip .1cm
 Dept. of Physics , Iran  University of Science and Technology,\\
Narmak, Tehran 16844, Iran.
\\ Institute for Studies in Theoretical Physics and 
Mathematics
 \\ Tehran P.O.Box: 19395-5746, Iran.
\end{center}
\vskip .5cm
\begin{abstract}
 
We calculate the coefficients of the operator product expansion (OPE), in   
Polyakov's approach for Burgers turbulence. We show that the OPE has
to be generelized and it is shown that the extra term gives us the instanton 
solution (shock solution) of Burgers equation. We consider the 
effect of the new-term in the OPE, on the right and left-tail of
probability distribution function (PDF). 
It is shown that the left-tail of PDF, where is dominated by 
the well-separated shocks behaves as $ W(u)\sim u^{-7/2}$.
Finally we calculate the assymptotic behaviour
of the N-point generating function of the velocity field, using the new OPE.
 
\end{abstract} 
\vskip .5cm

PACS numbers 47.27.AK, 47.27.Jv
\newpage
{\bf 1- Introduction}

A theoretical understanding of turbulence has eluded physicists for a 
long time. A statistical theory of turbulence has been put
forward by Kolmogorov [1], and further developed by others [2-4].
The approach is to model turbulence using stochastic
partial differential equations.  In this direction, Polyakov [5] has 
recently offered a field theoretic method to derive the 
probability distribution or density of states in (1+1)-dimensions 
in the problem of randomly driven Burgers equation. 
The importance of the Burgers equation is that, it is the simplest
equation that resembles the analytic structure of the Navier--Stokes equation, 
at least formally, 
within the scope of applicability of the Kolmogorov`s arguments [6].

Polyakov 
formulates a new method to analyse the inertial range correlation 
functions based on two important ingredients in field theory and 
statistical physics, the operator product expansion (OPE) and 
anomalies.
He argues that in the limit of high Reynolds number because of 
existence of singularities at the coinciding point, dissipation remains 
finite and 
all sublleading contributions vanish in the inertial range.
Using the OPE one can find the leading singularities and
 show that this approach is self-consistent.

Using the OPE Polyakov reduces the problem of computation of 
the veloctiy correlation functions to the 
solution of a certain partial differential equation.
In this direction the N-point generating functions have been found in [7]. The 
two-point functions for different type of correlation for noise has been found in [8] and 
the perturbative calculation in the presence of pressure has been done in [9].

Here using the results of [7], we calculate the OPE coefficients  
proposed by Polyakov, and show that we have to modify the OPE.
We write the generelized OPE and find the first correction to the PDF which 
was initially  found by Polyakov. It is shown that the extra term in the OPE
gives us the instanton, which has found by Gurarie, Migdal [10] and Bouchaud, Mezard [11].

We find the right and left-tail of the PDF and show that
the right-tail remaines unchanged and the left-tail ,
where is dominated by the shocks behaves as $ W(u)\sim u^{-7/2}$, which
is in agreement with recent numerical experiments [12].
Finally we find the assyptotic behaviour of the N-point generating 
function with the new-term (instanton-term) in the OPE.
\\
{ \bf {2-  The Instanton Solution of the Burgers Equation}} 
 
 We consider the Burgers equation in one-dimensin: 
\be
u_t + uu_x = \nu u_{xx} + f(x,t)
\ee
where  $u$ and $\nu $ are velocity field and viscosity, respectively.
The $ f(x,t)$ 
is a 
Gaussian random force with the following correlation:
\be
< f(x,t)  f(x^{'},t^{'})> =  k (x-x^{'}) \delta (t-t^{'}) 
\ee

To investigate the statistical description of eq.(1) following 
 Polyakov [5], we consider the following generating functional
\be 
Z_N(\lambda_1, \lambda_2,\ldots \lambda_N, x_1,\ldots x_N) = <\exp (
\sum^N_{j=1} \lambda_j u(x_j,t))>
\ee

Noting that the random force $ f(x,t)$ has a Gaussian distribution, 
$Z_N$ satisfies a closed differential 
equation provided that the viscosity  $\nu$ tends to zero:
\be
\dot{Z_N} + \sum \lambda_j {\partial \over {\partial \lambda_j}}({1\over 
{\lambda_j}}{\partial Z_N \over{\partial x_j}}) = \sum k(x_i-x_j) \lambda_i
\lambda_j Z_N +D_N
\ee
where $D_N$ is :
\be
D_N = \nu \sum \lambda_j <u^{''}(x_j,t) \exp \sum \lambda_k u(x_k,t)> 
\ee
The first term in the r.h.s. of eq.(4) can be derived by applying 
the following relatin and then using the Gaussian  
character of the noise term in time and also by giving address to causality. 
\be
<f(x,t) e^{\lambda u(x^{'},t)}> = \lambda \int^t dt^{'} < f(x,t) 
\dot { u} (x^{'}, t^{'}) e^{ \lambda u(x^{'},t^{'})}>
\ee

However to remain in the inertial range we must keep
$\nu$ infinitesimal but non-zero. Polyakov argues that the anomaly mechanism 
implies that infinitesimal viscosity produces a finite effect. To compute 
this effect, Polyakov makes a conjecture for existence of 
an operator 
product expansion or the fusion rules. The fusion rule is the statement 
concerning the behaviour of correlation functions, when some subset of 
points are put close together.

Let us use the following notation;
\be
Z(\lambda_1, \lambda_2,\ldots , x_1,\ldots x_N)= <e_{\lambda_1}(x_1)\ldots
e_{\lambda_N}(x_N)>
\ee
where $ e_{\lambda_1}(x_1) = e^{\lambda_1 u(x_1)}$.

The Polyakov`s conjecture is that in this case the OPE has the 
following form,
\be
e_{\lambda_1}(x+y/2) e_{\lambda_2}(x-y/2) = A(\lambda_1, \lambda_2, y)
e_{\lambda_1+\lambda_2}(x)+B(\lambda_1, \lambda_2, y){\partial \over \partial x}
e_{\lambda_1+\lambda_2}+o(y^2)
\ee

This implies that $Z_N$ fuses 
into functions $Z_{N-1}$ as we fuse a couple of points together. This 
conjecture
allows us to evaluate the following anomaly operator (i.e. the 
$D_N$-term in eq.(4)),
\be
a_\lambda(x)= \lim _{\nu\rightarrow 0}\nu (\lambda u^{''}(x) \exp 
(\lambda u(x))
\ee
which can be written as:
\be
a_\lambda(x)= \lim _{\xi,y,\nu\rightarrow 0}\lambda \nu {\partial^3\over 
{\partial \xi \partial y^2}} e_\xi (x+y) e_\lambda(x)
\ee
As discussed in [5] the possible Galilean invariant expression is:
\be
a_\lambda(x) = a(\lambda) e_\lambda(x) + \tilde{\beta}(\lambda){\partial
\over \partial x}e_\lambda(x)
\ee
Therefore in steady state the master equation takes the following form,
\bea
\sum ({\partial \over \partial \lambda_j} - \beta(\lambda_j)){\partial \over 
\partial x_j} Z_N -\sum \bar k(x_i -x_j) \lambda_i \lambda_j Z_N &=& \sum 
a(\lambda_j) Z_N \cr
\beta(\lambda) &=& \tilde{\beta}(\lambda)+{1 \over \lambda}
\eea

Let us consider following correlation for $ f$ in $k$-space as follows [13]:
\be
< f(k,t)  f(k^{'},t^{'})> =\frac {L}{2\pi}  k_0 \delta (k^2 - \frac{1}{L^2}) 
\delta (k+k^{'}) \delta (t-t^{'})        
\ee
therefore we obtain:
In the inertial range where $ x_i -x_j << L$, we find:
\be
k(x_i - x_j) =  k_0  (1- \frac{(x_i - x_j)^2}{2 L^2} ) 
\ee

Polyakov has found the following explicit form of $Z_2$ for 
$ k(x_i -x_j)$ given by eq.(14): 
\be
 Z_2(\mu y) = e^{\frac {2}{3}(\mu y)^{3/2}}
\ee
and the following expression for density of states as the Laplace transform
of $Z_2$ ;
\be                         
W(u,y) = \int_{c-i\infty}^{c+i\infty} {d\mu \over 2\pi i} e^{-\mu u} 
Z_2(\mu y)
\ee
where  $\mu=2 (\lambda_1 -\lambda_2)$ and $ y=x_1-x_2$. 
Now one can write the right and left--tail
of $W(u,y)$ as following:

\be
W(u,y) = \left \{ \begin {array} {ll}  e^{-\frac {1}{3} (\frac {u}{y})^3} &   
\mbox{if $(u/y)\rightarrow + \infty $}\\
 y^{3/2} u^{-5/2} + y^{9/2} u^{-11/2}&
 \mbox {if $(u/y)\rightarrow - \infty$}    
\end{array} \right.
\ee
In ref.[7] we have found the exact N-point generating function (i.e. the $Z_N$)
as follows:
\be
Z_N= (\lambda_1  \lambda_2 \cdots \lambda_N)^{b_N} (\mu_2 \mu_3 \cdots \mu_N)
^{-\frac{2N-1}{2(N-1)}} e^{2/3 (\mu_2 y_2 + \mu_3 y_3 +\cdots +\mu_N y_N)^{3/2}}  
\ee

where $b_N=\frac{2N-1}{2N}$ and $\mu_i$, $y_i$ are given by:

\bea
y_1&=&{{x_1 +x_2 +x_3 +\ldots x_N}\over N}\cr
y_2&=&x_1 -{{x_2 +x_3 +\ldots x_N}\over N-1}\cr
y_3&=&x_2 -{{x_3 +x_4 +\ldots x_N}\over N-2}\cr
and \hskip 1cm y_N&=& x_{N-1}-x_N 
\eea
and
\bea
\mu_1 &=&{{\lambda_1 + \lambda_2 +\ldots +\lambda_N}\over N}\cr
\mu_2 &=&{N-1\over N}[\lambda_1-{{\lambda_2 + \lambda_3 +\ldots +\lambda_N}
\over {N-1}}]\cr
\mu_3 &=&{N-2 \over{N-1}}[\lambda_2-{{\lambda_3 + \lambda_4 +\ldots 
+\lambda_N
}\over {N-2}}]\cr
and \hskip 1cm \mu_N&=& 2(\lambda_{N-1} - \lambda_N)
\eea
It follows that the $N$--point correlation function of $v$ is: 
\be
 G^{(N)} (x_1, \cdots ,x_N)\sim lim_{\lambda \rightarrow 0} \lambda^{-N} 
 \sum_{k=0} ^N a_k ^{(N)}   (\lambda x)^{\frac {3k}{2}} 
\ee
where $a_k^{(N)}$ are some constants.

Now we try to find the OPE coefficients in eq.(8).
To do this according [7] we use  $Z_3$ and tend $x_2$ close to $x_3$.
 $Z_2$ is given by eqs.(15) with $\lambda_1 + \lambda_2 =0$, and
 $Z_3$ has following form:
\bea
Z_3 &=& (\lambda_1 \lambda_2 \lambda_3)^{5/6} 
[3(\lambda_1 - \frac {\lambda_2 +\lambda_3}{2})
(\lambda_2 -\lambda_3)]^{-5/4} 
\nonumber \\  &.& e^{ \{2/3 [3/2 (\lambda_1 - \frac {\lambda_2 +\lambda_3}{2})
(x_1 - \frac {x_2 +x_3}{2})+2 (\lambda_2 -\lambda_3)(x_2-x_3)]^{3/2} \}}
\eea

where $\lambda_1 + \lambda_2 + \lambda_3 = 0$.
In eq.(22) we take $x_3 = x_2 - 2\epsilon$ and it is easy to show that in the limit
$\epsilon \rightarrow 0$ we find:

\bea
Z_3 &=& (\lambda_1 \lambda_2 \lambda_3)^{5/6}(9/2 (\lambda_3^2 - \lambda_2^2))^{-5/4}
e^{\{ 2/3[2(\lambda_1 - (\lambda_2 +\lambda_3))(x_1 -x_2)]^{3/2}\}} 
\nonumber \\ &\{& 1+ (-1 +7/{64} +\cdots ) 7/8 (\lambda_2 + \lambda_3) 
(-2) [-4 (\lambda_2+\lambda_3)]^{1/2} (x_1 -x_2)^{3/2}
\nonumber \\ &+& (1-7/{32} + \cdots ) \frac {7 \lambda_2 -25 \lambda_3}{32(\lambda_2+\lambda_3)}
(-2) [-4 (\lambda_2+\lambda_3)]^{3/2} (x_1-x_2)^{1/2} \epsilon + O(\epsilon^2) \}
\eea

Comparing with eq.(8) and using eq.(15) we find that:
\be
A(\lambda_2,\lambda_3,\epsilon) = (-\lambda_2 \lambda_3 (\lambda_2 + \lambda_3))^{5/6}
(\frac{9}{2} (\lambda_3^2 - \lambda_2^2))^{- \frac{5}{4}}
\ee
\be
B(\lambda_2, \lambda_3, \epsilon)=(1-7/32 +\cdots)  
\frac {7 \lambda_2 -25 \lambda_3}{32(\lambda_2+\lambda_3)}   
{A(\lambda_2\lambda_3,\epsilon)} \epsilon
\ee

Indeed we determine the $A(\lambda_2,\lambda_3,\epsilon)$ and $B(\lambda_2,\lambda_3,\epsilon)$
by means of the first and third terms in eq.(23).
However the second term in eq.(23) (i.e. which is porpotional to $(x_1 - x_2)^{3/2}$),
can be treated as $ \frac {\partial}{\partial \lambda} Z_2$ and this will change   
the OPE (i.e.  eq.(8)) to the following form:

\bea
e_{\lambda_2}(x+ \epsilon /2) e_{\lambda_3}(x- \epsilon /2)& =& A(\lambda_2, \lambda_3, \epsilon)
e_{\lambda_2+\lambda_3}(x)+B(\lambda_2, \lambda_3, \epsilon){\partial \over \partial x}
e_{\lambda_2+\lambda_3} \nonumber \\ &+& C(\lambda_2 , \lambda_3 , \epsilon)\frac {\partial 
}{\partial \lambda} e_{\lambda_2+\lambda_3} + O(\epsilon ^2) 
\eea
where using eq.(54), $C(\lambda_2, \lambda_3, \epsilon )$ can be written in terms of 
$A(\lambda_2, \lambda_3, \epsilon)$ as follows:
\be
C(\lambda_2 , \lambda_3 , \epsilon)=(-1 +7/{64} +\cdots) \frac{7}{8} 
(\lambda_2 +\lambda_3)  A(\lambda_2,\lambda_3,\epsilon)
\ee

Using eq.(26) one can show that the modified OPE leads to following equations for 
$a_\lambda (x)$ and $Z_2$ (i.e. eqs.(11) and (15)):
\be
a_\lambda(x) = a(\lambda) e_\lambda(x) + \tilde{\beta}(\lambda){\partial
\over \partial x}e_\lambda(x) + \tilde{\gamma}(\lambda){\partial
\over \partial \lambda}e_\lambda(x) +  \cdots
\ee
\be
(\partial_{\mu} - \frac {2b}{\mu})\partial_y Z_2 + c \mu \partial_\mu Z_2 -
\mu^2 y^2 Z_2=0
\ee
where we have used $\tilde{\beta}(\lambda)= \frac{b-1}{\lambda}$ , $a=0$ according
[5]. We have used the scaling arguments and show that $ \tilde{\gamma}(\lambda)$ has 
the following form:
\be
\tilde{\gamma}(\lambda)= c \lambda
\ee

Therefore we find following assymptotic behavior for $Z_2$:
\be
 Z_2(\mu y) = e^{\frac {2}{3}(\mu y)^{3/2} - \frac{c}{2} (\mu y)}
\ee

First of all it appears that in the limit $c\rightarrow 0$, one finds the Polyakov`s
result for $Z_2$. Also by using of eq.(28), one can show that the 1--point 
function of the velocity behaves as $<v(y)> \sim y $, which is 
related to the well-known shock wave structures [6], and is 
exactly the instanton solution of the noisy Burgers equation, found 
by Gurarie and Migdal [10], and later in a different way by Bouchaud and Mezard [11]. 
The interesting point 
is that Polyakov`s approach generates in self-consistent  way the instanton solution.

Let us write the right and the left-tail of the PDF. 
It is easy to show that the $W(u,y)$ behaves as follows:

\be
W(u,y) = \left \{ \begin {array} {ll}  e^{-\frac {1}{3} (\frac {u}{y}+\frac {c}{2})^3} &   
\mbox{if $(u/y)\rightarrow + \infty $}\\
 y^{3/2} u^{-5/2} + c y^{5/2} u^{-7/2} + c^2 y^{7/2} u^{-9/2} + y^{9/2} u^{-11/2}&
 \mbox {if $(u/y)\rightarrow - \infty$}    
\end{array} \right.
\ee

It is evident that the right-tail remanies somewhat unchanged and the left-tail 
of the PDF, where is dominated by the shocks behaves as $W(u)\sim u^{-7/2} $,
which is in agreement with the observations [12].

Now let us find the assymptotic behaviour of the N-point generating functions.
For investigating the assymptotic behaviour we use the  
 following important property that:
\be
\Sigma_{i=1} ^N \lambda_i \partial_{\lambda_i} = 
\Sigma_{i=1} ^N \mu_i \partial_{\mu_i} 
\ee

where $\mu_1=0$, which in turn gives the following 
assymptotic behaviour of the N-point generating functions 
in the limit of $\Sigma_{i=1} ^N \mu_i y_i \Rightarrow \infty$:

\be
Z_N= (\lambda_1  \lambda_2 \cdots \lambda_N)^{b_N} (\mu_2 \mu_3 \cdots \mu_N)
^{-\frac{2N-1}{2(N-1)}} e^{2/3 (\mu_2 y_2 + \mu_3 y_3 +\cdots +\mu_N y_N)^{3/2}}  
e^{-c/2 \Sigma_{i=1} ^N \mu_i {y_i}}
\ee

Now it is easy to show that for un-forced Burgers equation ($k_0=0$), we find,
\be
<u^q (r)> \sim r ^q
\ee
which is again cosistent with other approaches [14].
\\
{\bf Acknowledgements:} I would like to thank A. Rastegar, S. Rouhani and 
J. Davoudi for valuable 
discussions. 

\end{document}